\documentclass[groupaddress]{revtex4}
\usepackage{amsthm}
\usepackage{graphicx}  
\usepackage{dcolumn}   
\usepackage{bm}        
\usepackage{amssymb}   
\usepackage{amsmath}
\usepackage{epsfig}
\usepackage{rotating}
\usepackage{xspace}
\usepackage{hyperref}
\usepackage{xcolor}

\newcommand{\C}{\mathbb{C}}
\newcommand{\Z}{\mathbb{Z}}
\newcommand{\R}{\mathbb{R}}

\newcommand{\cO}{{\cal O}}

\newcommand{\cV}{{\cal V}}

\newcommand{\e}{\text{e}\,}
\newcommand{\bra}{\langle}
\newcommand{\ket}{\rangle}

\newcommand{\no}{\noindent}
\newcommand{\bear}{\begin{eqnarray}}
\newcommand{\eear}{\end{eqnarray}}

\def\be#1\ee{\begin{equation}#1\end{equation}}
\def\bea#1\eea{\begin{align}#1\end{align}}

\begin{document}

\title{
Complex Langevin: Boundary terms at poles
}

\author{Erhard Seiler}
\email{ehs@mpp.mpg.de}
\affiliation{Max-Planck-Institut f\"ur Physik (Werner-Heisenberg-Institut),  
F\"ohringer Ring 6, M{\"u}nchen, Germany}

\date{\today}

\begin{abstract} 
We discuss the problem of possible boundary terms at poles of the drift in 
the complex Langevin method, which spoil correctness of the method. For 
the simplest, however paradigmatic cases we can find complete answers. 
Lessons for more generic cases as well as open mathematical problems are 
discussed.
\end{abstract}
\maketitle




\section{Introduction}

The complex Langevin (CL) method has been studied for almost 40 years 
\cite{parisi, klauder}, but it still has unsolved mathematical aspects. 
We have given a formal justification of the method in \cite{aarts1, 
aarts2}, pointing out already the possible failure of the justification 
due to unwanted boundary terms ``at infinity''. In an interesting paper, 
which among other results presents some new criteria for the possible 
failure of the CL method, Salcedo \cite{salc} provided a different 
perspective on these boundary terms.

In \cite{scherzer1} and \cite{scherzer2} we studied these boundary terms 
in great detail for some models, leading to verifiable criteria for 
correctness and even to the computation of corrections in cases of 
failure. Since we were dealing with holomorphic drifts, all these 
boundary terms were at infinity, resulting from slow decay of the 
probability, since we were dealing with holomorphic drifts.

A particularly thorny issue is the problem of meromorphic drift arising 
from zeroes of the complex density defining the models, a problem that 
is unavoidable in finite density QCD due to zeroes of the fermion 
determinant. This makes it necessary to consider also boundary terms at 
the singularities of the drift. Nishimura and Shimasaki studied this 
problem for simple models \cite{nishi}; Aarts {\it et al.} \cite{pole} later 
presented a detailed study of this issue, with the emphasis on numerical 
analysis of various models from the simplest one-dimensional case to full
QCD.

In this paper I am beginning a mathematical analysis of the boundary 
terms at poles of the drift, focusing on the simplest model already 
studied in \cite{pole}, the so-called one-pole model. In fact, I start 
with a further simplification of that model in which only the pole term 
of the drift is kept; this allows to clarify the issue by carrying out 
explicit calculations. A heuristic justification for this simplification 
is the realization that near the pole, this term dominates the drift, so 
neglecting the rest of the drift should be a good approximation of the 
CL process while it is spending time in the neighborhood of the poles. 
In this simple approximation the linkage of the failure of the CL method 
with the appearance of boundary terms at the poles becomes manifest.

In \cite{pole} it was incorrectly claimed that those boundary terms had 
been found in the long time equilibrium limit; this error was pointed 
out by Salcedo \cite{salcedo}; see the erratum to \cite{pole}. So 
here we move away from equilibrium and consider the short time 
evolutions, where we do find indeed the sought for boundary terms.

For the benefit of the reader we briefly recapitulate the general idea 
of the formal justification of the CL method and show where boundary 
terms may arise at poles of the drift, invalidating the formal 
justification.
 
I consider a complex density on $\R$
\be
\rho(x)=\exp(-S(x))\,,\quad \int dx \rho(x)=1\,,
\ee
where $\rho$ extends to an entire analytic function, but with possible zeroes 
(in which case $S$ is of course multivalued). 

\no The complex Langevin equation (CLE) in the form used here is
\begin{align}
dx=&K_x dt +dw,\notag\\ dy=&K_y dt\,,
\label{cle2}
\end{align}
where $dw$ is the Wiener process normalized as \be \bra dw^2\ket= 2 dt\,
\ee
and the drift is given by
\begin{align}
&K=-S'=\frac{\rho'}{\rho}; \cr &K_x= {\rm Re}K\,\quad  K_y= {\rm Im}K\,.
\end{align}
The drift thus is univalued but has simple poles at the zeroes of $\rho$.
The average of a generic holomorphic observable $\cO$ is denoted by
\be
\bra \cO \ket_{t,z_0}\equiv\int dx dy P_{z_0}(x,y;t)\cO(x+iy)\,;  
\ee
where $P_{z_0}$ is the probability density on $\C$ produced by the CL 
process starting at $z_0=x_0+iy_0$ and running for time $t$; the time 
evolution of $P$ is given by the Fokker-Planck equation (FPE). We say 
that the CL process yields correct results if $\bra \cO 
\ket_{\infty,z_0}$ agrees with the ``correct'' expectation value of the 
same observable defined as
\be
\bra {\cO} \ket_c = \int dx {\cO}(x) \rho(x)\,,
\ee
i.~e. if
\be
\bra {\cO} \ket_{\infty,z_0}= \bra {\cO} \ket_c\,.
\label{correctness}
\ee
In \cite{aarts1,aarts2} it was shown that correctness is assured if the
so-called interpolating function 
\be
F_\cO(t,\tau)= \int_0^\infty dx P_{z_0}(x;t-\tau) \cO(x;\tau)
\label{interpolfct}
\ee
is independent of the parameter $\tau\in[0,t]$. Here $\cO(z;\tau)$ is the 
solution of the initial value problem
\be
\frac{\partial}{\partial \tau}\cO(z;\tau)=L_c\cO(z;\tau)\,,\quad 
\cO(z;0)=\cO(z)\,;\quad L_c=(\partial_z+K(z))\partial_z\,,
\label{cauchy}
\ee
The interpolation property follows from 
\be
F_\cO(t,0)=\bra \cO(x_0+iy_0)\ket_{t,x_0}\,;\quad 
F_\cO(t,t)= \cO(x_0+iy_0;t)\,, 
\ee
so 
\be
\partial_\tau F_\cO(t,\tau)=0 \Longrightarrow
\bra \cO\ket_{t,z_0}= \bra\cO(z_0;t)\ket_0 \quad \forall t > 0\,,
\label{interpol}
\ee
from which correctness (\ref{correctness}) can be deduced. The left-hand side 
of (\ref{interpol}), via integration by parts, is equal to a boundary term, 
arising from possible slow decay at infinity as well as from poles of the 
drift. Details are found for instance in \cite{aarts1,pole}.

\section{The need to consider the evolution before reaching equilibrium}

In \cite {scherzer1,scherzer2} we found boundary terms at infinity by 
considering the equilibrium distributions. But we could not find 
boundary terms near poles that way, because the equilibrium distribution 
$P(x,y;t=\infty)$ of the probability density was always found to vanish 
at least linearly at the poles of the drift, so holomorphic observables 
could not lead to boundary terms at the pole, as we will see.

The argument goes as follows: for simplicity let us assume that there is 
a pole at the origin; for the boundary term arising in equilibrium and for 
$\tau=0$ (see \cite{pole}) consider
\be
\int_{x^2+y^2\le\delta^2} dx\,dy P(x,y;t=\infty) 
L_c\cO(x+iy)\,.
\label{volume}
\ee
Using integration by parts and the Cauchy-Riemann equations (\ref{volume}) 
is
\be 
\int_{x^2+y^2\le\delta^2}  dx\,dy \cO(x+iy) (L^TP)(x,y;t=\infty) 
+B_{\delta}= B_{\delta}\,
\label{boundterm}
\ee
(where $L^T$ is the Fokker-Planck operator, see \cite{aarts1}), since 
the first term of the left-hand side vanishes in equilibrium. 
$B_{\delta}$ is a boundary term. Now, since $\cO$ is holomorphic, 
$L_c\cO$ has at most a simple pole at the origin, stemming from the 
pole in the drift. Since $P$ vanishes linearly at the origin, the 
integrand of (\ref{volume}) is bounded in the region of integration, 
hence the boundary term vanishes for $\delta\to 0$. If we consider the 
time evolution for finite time $t$, the boundary term now is 
given by
\be
B_\delta=\int_{x^2+y^2\le\delta^2} dx\,dy \left\{\cO(x+iy)L^T 
P_{z_0}(x,y;t)-P_{z_0}(x,y;t) L_c\cO(x+iy)\right\}\,
\label{full_bt}
\ee
and the first term of this expression is no longer zero. In fact, below we 
will give an example where the second term of (\ref{full_bt}) vanishes, but 
there is a boundary term arising solely from the first term of 
(\ref{full_bt}).

The main difficulty is now to understand the $L_c$ evolution 
(\ref{cauchy}) of observables  in the presence of poles. 
We focus on the simplest model, dubbed one-pole model in \cite{pole}. 
Since the equilibrium distribution does not lead to a boundary term, in 
this note we focus on the short time evolution, and we do indeed find 
boundary terms there.

\section{The one-pole model}

The action for the one-pole model can be written (after shifting the
contour of integration) as
\be
S=-\ln\rho(x) = -n_p \ln x +\beta (x+z_p)^2\,,
\ee
\be
\rho(x)= x^{n_p} \exp(-\beta (x+z_p)^2)\,
\ee
with $n_p$ a positive integer. The drift of the CL process is then given by 
the real and imaginary parts of
\be
K(z)=\frac{\rho'}{\rho}=\frac{n_p}{z}-2\beta (z+z_p)\,
\ee
and the complex Langevin operator determining the evolution of holomorphic 
observables is
\be
L_c=(D_z+K)D_z=(D_z+\frac{n_p}{z}-2\beta (z+z_p))
D_z\,.
\ee
where we wrote
\be
D_z \;\; {\rm for}\;\; \frac{d}{dz}\,
\ee
and we later use the same symbol for the partial derivative.

\section{The ``pure pole model'': $\beta=0$}

Since we are not analyzing the equilibrium distribution we have the 
freedom to study systems that do not possess one; this leads to the 
consideration of the one-pole model for $\beta=0$. This is the 
absolutely simplest model having a pole in the drift. Since $z_p$ plays 
no role, we also set it equal to zero.

We compare the finite time evolution of the probability density under 
the CL process with the evolution of the observables under 
(\ref{cauchy}) or equivalently the semigroup $\exp(tL_c)$. The two 
evolutions should be consistent if there are no boundary terms (see 
\cite{aarts1, aarts2, pole}).

For $\beta=0$ the system could be treated by a real Langevin process. 
We will nevertheless study the system in the complex domain by choosing 
a complex starting point for the Langevin process.

\subsection{$n_p=2$}

\subsubsection{The $L_c$ evolution}

The Langevin operator for this special case is
\be
L_c=D_z^2+\frac{2}{z} D_z\,.
\ee
For $n_p=2$ there is a simplification, pointed out already in 
\cite{pole}: $L_c$ is related to the Langevin operator with zero drift 
by a similarity transformation:
\be
L_c= \frac{1}{z}\,D_z^2\, z
\ee
and hence   
\be
\exp(tL_c)=\frac{1}{z} \exp(t D_z^2) z\,,
\ee
with the integral kernel
\be
\exp(tL)(z,x')= \frac{x'}{z} \frac{1}{\sqrt{4\pi t}}
\exp\left(-\frac{(z-x')^2}{4t}\right)\,.
\ee
Here $x'$ is to be understood as an integration variable along the real 
axis. The evolution of a holomorphic observable $\cO(z)$ is thus given by
\be
\cO(z;t)=
\frac{1}{\sqrt{4\pi t} z}\int_{-\infty}^{\infty} dx' x'
\exp\left(-\frac{(z-x')^2}{4t}\right) \cO(x')\,.
\ee
For the observables $\cO_k(z)\equiv z^k, k=1,\ldots 4$ and $k=-1$ this yields
\be
\cO_1(z;t)= z+\frac{2t}{z}\,,
\label{z}
\ee
\be
\cO_2(z;t)=z^2+6t\,,
\label{z^2}
\ee
\be
\cO_3(z;t)=z^3+12 t z+\frac{12t^2}{z}\,,
\label{z^3}
\ee
\be
\cO_4(z;t)=z^4+20 t z^2+60t^2
\label{z^4}
\ee
and
\be
\cO_{-1}(z;t)=\frac{1}{z}\,.
\label{1/zfull}
\ee
More generally, it is easy to prove inductively that for any $k\ge -1$ 
$\cO_k(z;t)$ is a polynomial in $t$; for even $k$ it is holomorphic, 
while for odd $k$ it is meromorphic with a simple pole at $z=0$. So 
$\exp(tL_c)$ applied to polynomial observables is indeed given by the 
exponential series, which actually terminates after finitely many terms.

\subsubsection{Comparison with the FPE evolution}

To study the FPE evolution, we have to resort to numerical simulation. 
We proceed by running 10000 trajectories of the CL process all with a 
fixed starting point $z_0$, stopping after Langevin times $t=0.01,\,0.1,\,
0.5,\,2.0$. 

Comparing the FPE results with those of the $L_c$ evolution, we find 
there is good agreement for the even powers but drastic disagreement for 
the odd ones, except for very small times.

\begin{figure}[ht]
\begin{center}
\includegraphics[width=0.48\columnwidth]{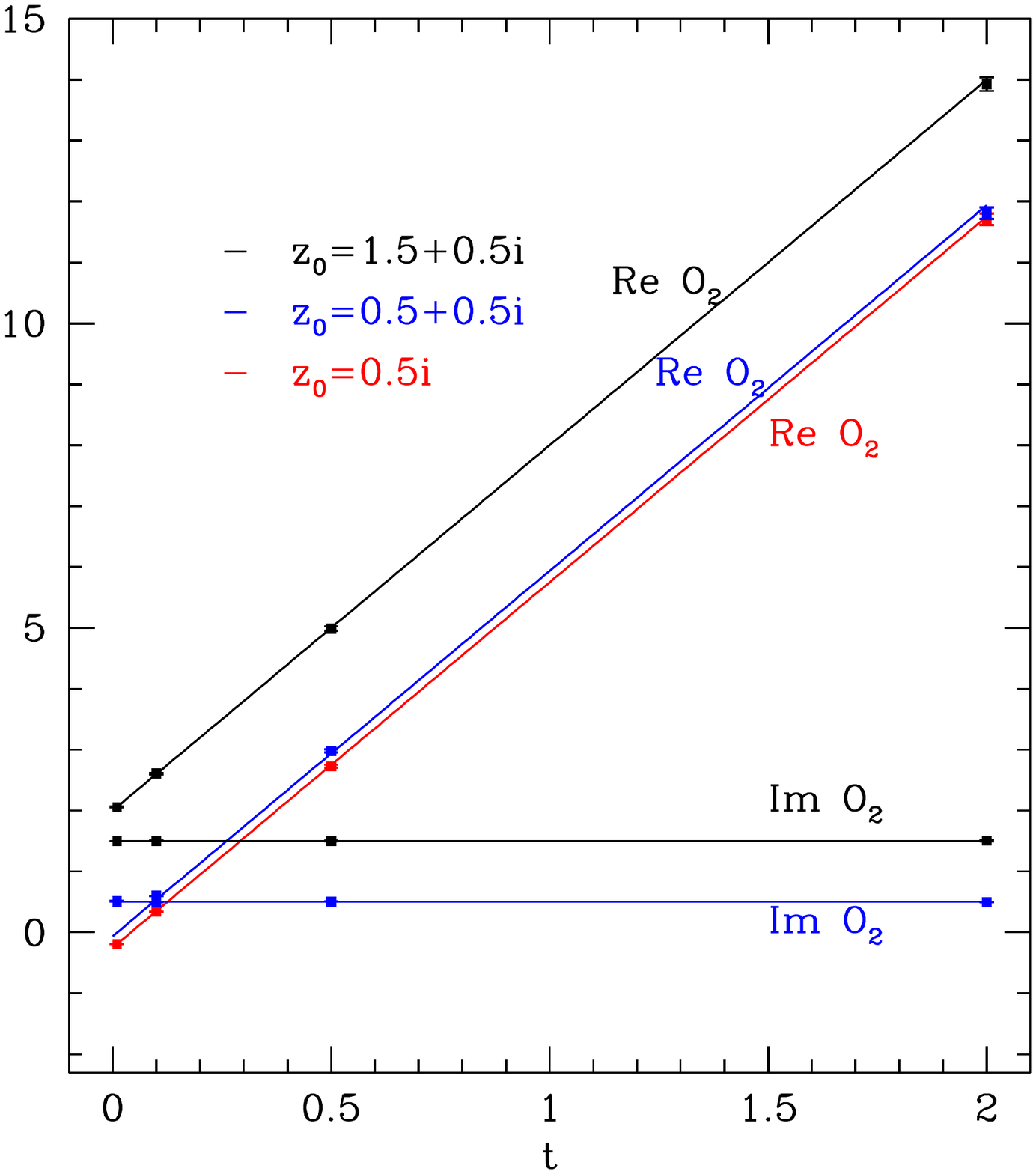}
\includegraphics[width=0.48\columnwidth]{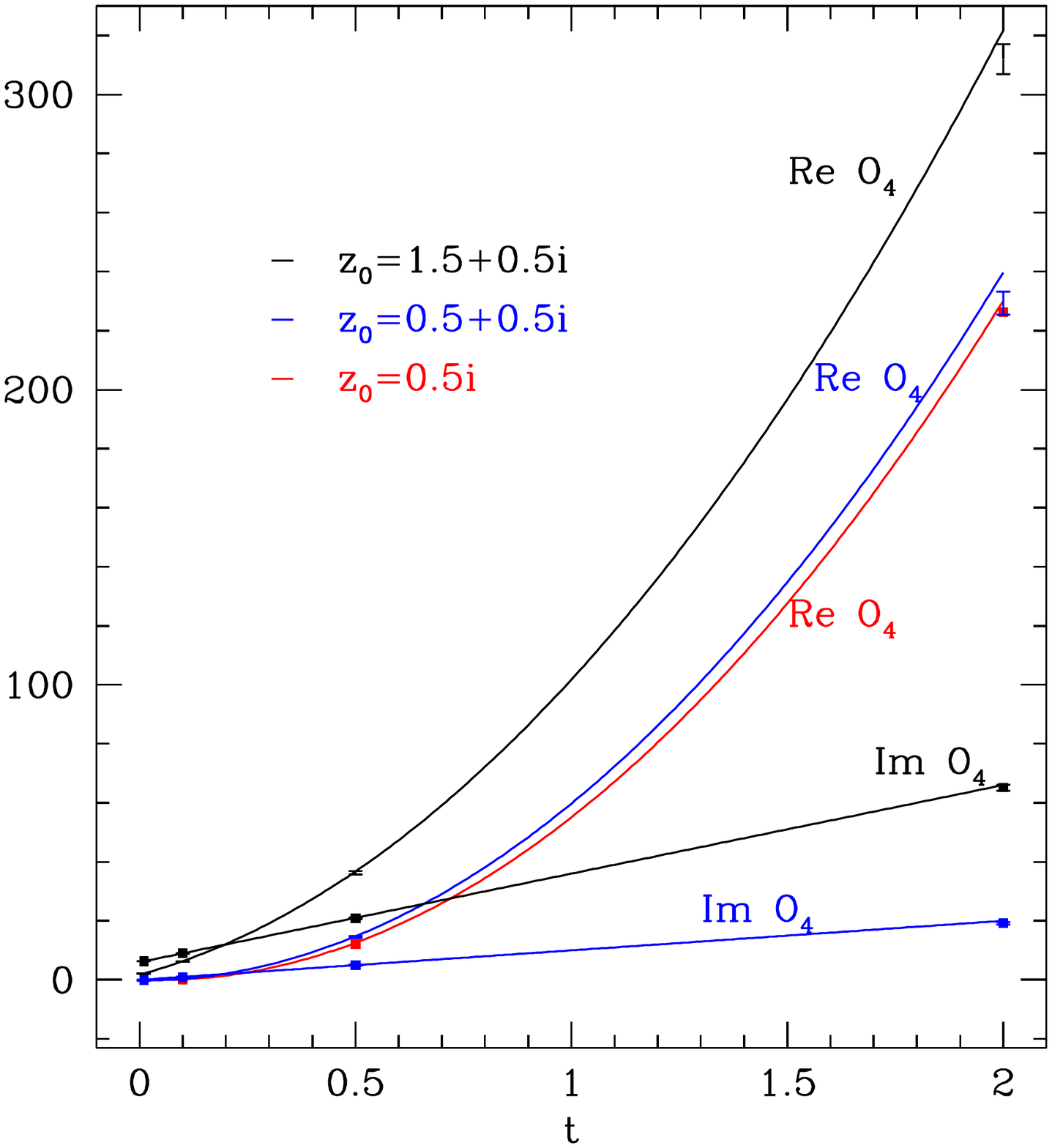}
\vglue-2cm
\caption{Comparison of $\cO_k(z;t)$ with $\bra \cO_k \ket_{t,z_0}$ for $k=2$ 
(left) and $k=4$ (right) for $\beta=0$.}
\label{fpecomp}
\end{center}
\end{figure}

In Fig.\ref{fpecomp} we show two plots comparing the evolution of the 
even powers $\cO_2(z_0;t)$ and $\cO_4(z_0;t)$ with the corresponding 
results $\bra \cO_2\ket_{t,z_0}$ and $\bra \cO_4\ket_{t,z_0}$ based on the 
FPE, for starting points $z_0=0.5i$, $z_0=0.5+0.5i$ and $z_0=1.5+0.5i$.

The agreement between the FPE and $L_c$ evolutions for even powers
corresponds to the fact that in this case there are no $1/z$ terms
appearing, hence no boundary terms at the origin.

\begin{figure}[ht]
\begin{center}
\includegraphics[width=0.6\columnwidth]{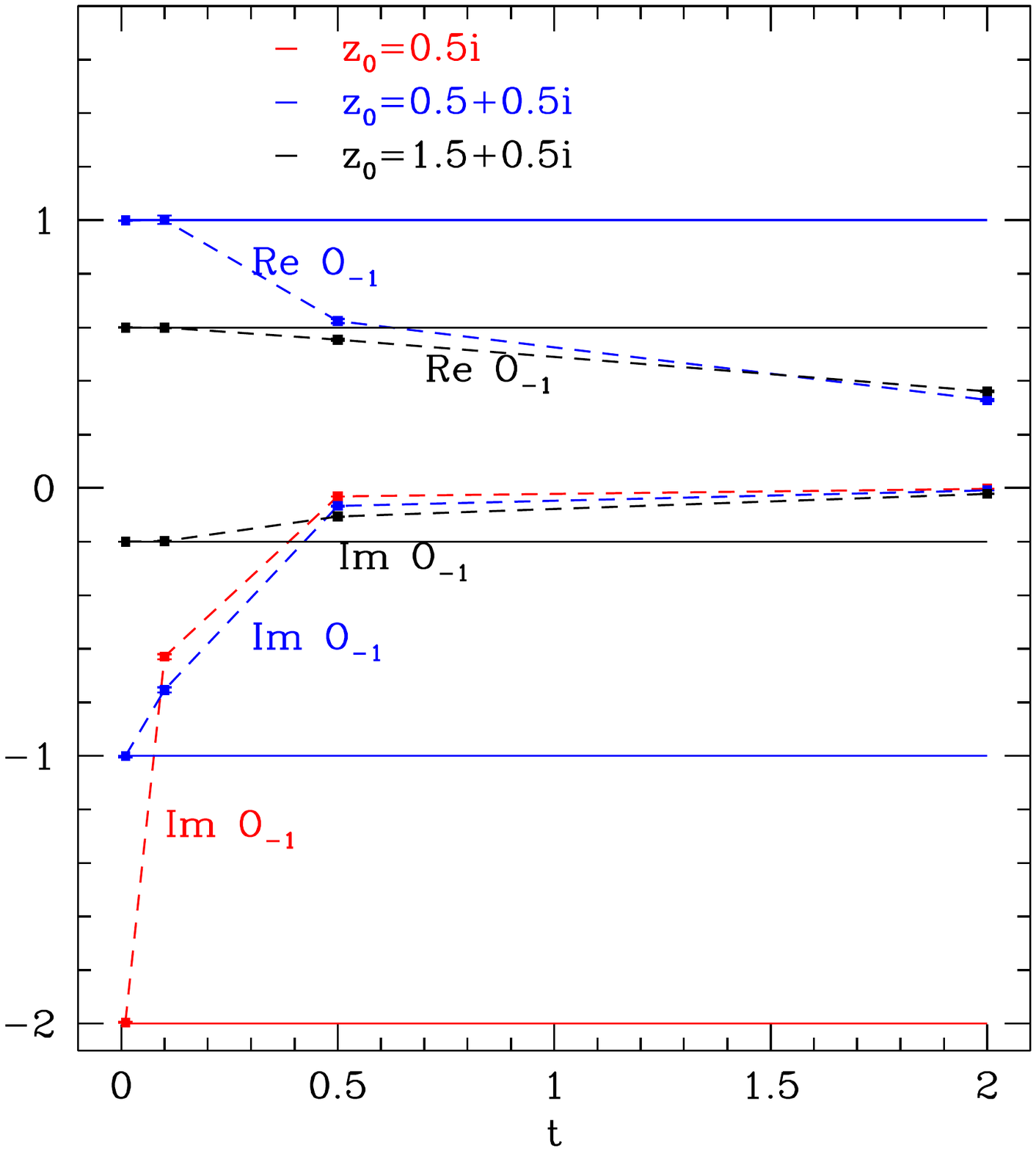}
\vglue-2cm
\caption{Comparison of $\cO_k(z_0;t)$ with $\bra \cO_k\ket_{t,z_0}$ for $k=-1$  
and $\beta=0$. Solid lines: correct results, dashed lines: CL results,
connected to guide the eye.}
\label{fpe0comp1m}
\end{center}
\end{figure}

The opposite is true for the odd powers, there is strong disagreement, 
indicating the presence of a boundary term. As an example we show in 
Fig.\ref{fpe0comp1m} the comparison of the FPE and $L_c$ evolutions for 
the observable $\cO_{-1}$ and the same three starting points 
$z_0=0.5i$, $z_0=0.5+0.5i$ and $z_0=1.5+0.5i$; some data of the 
comparison are compiled in Table \ref{table1}.

\begin{table}[ht]
\begin{center}
\begin{tabular}{|r||r |r| r| r|}
\hline
$t$ & 0.01 & 0.1 &  0.5 & 2.0 \\
\hline\hline
$\bra \cO_1 \ket_{t,z_0}$ & 0.45982(2)i & 0.1564(10)i& 0.05909(66)i& 0.0278(41)i \\
\hline
$\cO_1(z;t)$ & 0.46I &0.1i & -1.5i & -7.5i \\
\hline\hline
$\bra \cO_2 \ket_{t,z_0}$ &-0.18906(3) &0.3435(35)&2.727(21)& 12.0794(10) \\
\hline
$\cO_2(z;t)$ & -0.19 & 0.35 & 2.75 & 11.75  \\
\hline\hline
$\bra \cO_3 \ket_{t,z_0}$ & -0.06623(43)i&0.2356(34)i&0.650(11)i &1.284(27)i\\
\hline
$\cO_3(z;t)$ &-0.0674i &0.235i &-3.125i & -84.125i \\
\hline\hline 
$\bra \cO_4 \ket_{t,z_0}$ &0.01756(36)& 0.1550(46)&12.08(20)&226.3(3.9)\\
\hline
$\cO_4(z;t)$& 0.0185 & 0.1625 &12.5625 & 230.063  \\
\hline\hline
$\bra \cO_{-1} \ket_{t,z_0}$ &-1.9957(21)i&-0.6297(86)i&-0.03144(58)i
&-0.00356(8) i\\
\hline
$\cO_{-1}(z;t)$ & -2i &-2i &-2i & -2i  \\
\hline\hline
\end{tabular}
\vglue5mm
\caption{Comparison of CL results with $\cO_k(z;t)$ for $z_0=0.5i$, $n_p=2$.}
\label{table1}
\end{center}
\end{table}

\subsubsection{Interpolating function and boundary term}

From the failure of agreement for the odd powers it is easy to see that
the interpolating function $F_{\cO_k}(t,\tau)$ is {\em not} independent of
$\tau$ when $k$ is odd. We choose as the simplest case
$\cO_{-1}=1/z$; according to (\ref{1/zfull}) $\cO_{-1}(z;\tau)=1/z$, 
independent of $\tau$, so according to (\ref{interpolfct})

\be
F_{\cO_{-1}}(t,\tau)=\int P_{z_0}(x,y;t-\tau)\frac{1}{x+iy} dx dy\,.
\ee
From Fig.\ref{fpe0comp1m} and Table 1 it is clear that this is {\em not} 
independent of $\tau$, and
\be
\frac{d}{d\tau}F_{\cO_{-1}}(t,\tau)=\int (L^T P_{z_0})(x,y;t-\tau)
\frac{1}{x+iy} dxdy\neq 0\,.
\ee
As discussed in Sec.~2, this is a boundary term, and it can only be 
due to the pole at $z=0$, because for finite time $P_{z_0}$ shows strong 
(Gaussian) decay. It could be evaluated also directly as a boundary term, 
but this is not necessary.

We can also see that there is no boundary term for even observables, 
such as $\cO_2(z)=z^2$. While it is difficult to evaluate the $\tau$ 
derivative directly because it involves $L^T P(x,y:t-\tau)$, we can 
compute the interpolating function for different values of $\tau$ to 
see that it is constant:

Let us take for instance $t=2.$ and $t-\tau=0.01,\;0.1,\;0.5,\;2.0$. We 
then find, using (\ref{z^2},\ref{z^4}) 
\bea F_{\cO_2}(2,\tau)&=6\tau+\bra \cO_2 \ket_{t-\tau,z_0}
\notag\\ F_{\cO_4}(2,\tau)&=60 \tau^2+20\tau\bra \cO_2
\ket_{t-\tau,z_0}+\bra \cO_4 \ket_{t-\tau,z_0}\,, 
\label{interpolval} 
\eea 

In Table 2 we present the values of these quantities, showing independence 
of $\tau$ within the errors.
 
\begin{table}[ht]
\begin{center}
\begin{tabular}{r}
 $z_0=0.5i$:
\end{tabular}
\begin{tabular}{|r||r| r| r| r| r|}
\hline
$\tau$ &2.0 &1.99 & 1.9& 1.5 & 0.0 \\
\hline
\hline
$F_{\cO_2}(2,\tau)$& 11.75(0) & $11.7509(3) $ & $11.7435(35) $ & $11.727(22)
$
& $11.759(98) $
\\
\hline
$F_{\cO_4}(2,\tau)$& 230.0625(0) & $230.099(12) $ & $229.81(13) $ &
$228.89(66)$ & $234.3 (4.4)$\\
\hline
\end{tabular}
\vglue5mm
\begin{tabular}{r}
 $z_0=0.5+0.5i$:
\end{tabular}
\begin{tabular}{|r||r| r| r| r| r|}
\hline
$\tau$ &2.0 &1.99 & 1.9& 1.5 & 0.0 \\
\hline
\hline
Re$F_{\cO_2}(2,\tau)$& 12.0(0) & $12.0009(15) $ & $12.0000(61)$ &
$11.981(25) $& $11.978 (99)$\\
\hline
Im$F_{\cO_2}(2,\tau)$& 0.5(0) & $0.5002(14) $ & $0.4998(37)$ &
$0.5046(54)$& $0.5016(71) $\\
\hline
Re $F_{\cO_4}(2,\tau)$&239.75(0) & $239.788(60) $ & $239.75(23) $ &
$238.96(75)$ & $ 239.9(4.4)$\\
\hline
Im $F_{\cO_4}(2,\tau)$& 20.0(0) & $20.003(56) $ & $19.99(14) $ &
$20.12(16)$ & $20.27(49)$\\
\hline
\end{tabular}
\caption{The interpolating functions (\ref{interpolval}) for various values
of $\tau$ and starting points $z_0=0.5i$ and $z_0=0.5+0.5i$.}
\end{center}
\label{table2}
\end{table}

\subsection{Remarks on general $n_p>0$}

We have
\be
L_c=D_z^2+\frac{n_p}{z}D_z\,.
\ee
This operator still leaves the linear space spanned by the even non-negative
powers $z^{2\ell}$,  $\ell\ge 0$ invariant, for any integer $n_p>0$. 
The odd powers $z^{2\ell-1}$ for $\ell\ge n_p/2$ span an invariant linear 
space as well.

But for $n_p$ odd, iterating the application of $L_c$ to $\cO_k$ will not 
terminate. For $n_p$ even, however, it does terminate at the power 
$z^{1-n_p}$; the observable 
\be 
\cO_{1-n_p}(z)\equiv z^{1-n_p} 
\ee 
is an eigenvector with eigenvalue $0$. As for $n_p=2$, for $k$ even 
$\bra \cO_k\ket_{t,z_0}$ and $\cO_k(z_0;t)$ agree , whereas for the odd 
powers they disagree. Likewise we find that $\exp(tL_c)\cO_k$ is a 
polynomial in $t$; for $k=2\ell$ it will be a polynomial in $z$, whereas 
for $k=2\ell-1$ it will be a rational function of $z$ with the lowest 
negative power being $z^{1-n_p}$.

As an example, we consider three observables for the case $n_p=4$:
we find by a simple calculation 
\be
\cO_2(z;t)=z^2+2(1+n_p)t\,,
\label{np4_z^2}
\ee
\be
\cO_4(z;t)=z^4+4(3+n_p)tz^2+4(1+n_p)(3+n_p)t^2\,,
\label{np4_z^4}
\ee
\be
\cO_{1-n_p}(z;t)=z^{1-n_p}\,.
\label{np4_z^{-}}
\ee


Some data comparing $\cO_k(z;t)$ with $\bra \cO_k \ket_{t,z_0}$  for the case 
$n_p=4$ are compiled in Table 3.

\begin{table}[ht]
\begin{center}
\begin{tabular}{|r||r |r| r| r| r|}
\hline
$t$ & 0.01 & 0.1 &  0.5 & 1.0 & 2.0 \\
\hline\hline
$\bra \cO_2 \ket_{t,z_0}$ &-0.14981(30)&0.7528(50)&4.731(30)&9.776(61)&
19.56(12)\\   
\hline
$\cO_2(z;t)$ & -0.15 & 0.75 & 4.75 & 9.75 & 19.56 \\
\hline\hline
$\bra \cO_4 \ket_{t,z_0}$ &0.0660(30) & 0.768(11)&31.03(41)&132.7(1.7)&529.5(6.9)\\
\hline
$\cO_4(z;t)$ & -0.065 & 0.7625 & 31.5625  & 133.063 & 546.063 \\
\hline\hline
$\bra \cO_3 \ket_{,z_0}t$ &11.458(027 i)&0.384(18)i&0.384(18)i&0.00078(7)i
&0.00013(2)i\\
\hline
$\cO_{-3}(z;t)$ & 8i & 8i & 8i & 8i & 8i \\
\hline\hline

\end{tabular}
\vglue5mm
\caption{Comparison of CL results with $\cO_k(z;t)$ for $z_0=0.5i$, $n_p=4$.}
\label{table3}
\end{center}
\end{table}

\section{$\beta>0$ and $z_p=0$}

\subsection{$n_p=2$}

The Langevin operator is now
\be
L_c=D_z^2+\left(\frac{2}{z}-2\beta z\right) D_z
\ee
with
\be
D_z=\frac{d}{dz}.
\ee
As pointed out in \cite{pole}, the similarity transformation,
\be
\exp(-S/2)\,L_c\, \exp(S/2) \equiv -H_{FP}=
(D_z+\frac{1}{2}K)(D_z-\frac{1}{2}K)\,
\label{simil1}
\ee
after restricting to the real axis yields in this case essentially the
Hamiltonian of a harmonic oscillator
\be
H_{FP}=-D_x^2+\beta^2 x^2-3\beta \,.
\ee
Equation (\ref{simil1}) implies the relation for the semigroups
\be
\exp(-S/2)\,\exp(tL_c)\, \exp(S/2)=\exp(-tH_{FP}).
\label{simil2}
\ee
Note that $H_{FP}$ is {\em not} positive on $L^2(\R)$; it has exactly one 
negative eigenmode:
\be
\psi_0(x)\propto\exp\left(-\frac{\beta}{2} x^2\right)
\ee
In \cite{pole} it is explained that this problem disappears if one 
considers $H_{FP}$ on $L^2(\R_+)$ with Dirichlet boundary conditions at 
$0$; then only the odd eigenvectors contribute. Because $\rho(x)$ 
vanishes at $x=0$ the (real) Langevin process avoids the origin.

In any case, because of (\ref{simil2}) we can use Mehler's formula
\cite{barry}
\bea
\exp(-t H_{FP})(x,y)=&-\sqrt{\frac{\beta}{\pi (1-e^{-2\beta
t})}}\notag\\ \times
&\exp\left[-\frac{\beta (x^2+y^2)}{2\tanh(2\beta t)}\right]
\exp\left(\frac{\beta x y}{\sinh(2\beta t)}\right)\exp(2\beta t)\,.
\eea
to obtain the kernel for $\exp(tL_c)$:
\bea
\exp(tL_c)(x,y)
=&\frac{y}{x}\exp\left(\frac{\beta}{2}(x^2-y^2)\right)\exp(2\beta t)
\sqrt{\frac{\beta}{\pi (1-e^{-4\beta t})}}\notag\\ \times
&\exp\left[-\frac{\beta (x^2+{y}^2)}{2\tanh(2\beta t)}\right]
\exp\left(\frac{\beta  x y}{\sinh(2\beta t)}\right)\,.
\label{lc_semigroup_full}
\eea
We define
\be
b\equiv\frac{\beta}{\sinh(2\beta t)}\,;\quad
\sigma\equiv\frac{1}{\beta(\coth(2\beta t)+1)}=
\frac{1-\exp(-4\beta t)}{2\beta}\,,
\label{params}
\ee
so that
\be
b\sigma=\exp(-2\beta t)\,.
\ee
Thus (\ref{lc_semigroup_full}) becomes
\be
\exp(tL_c)(x,y)=A(x;t)y \exp\left(-\frac{y^2}{2\sigma}\right)\exp(bxy)
\ee
with 
\be
A(x;t)=\frac{1}{x} \sqrt{\frac{\beta}{\pi (1-e^{-4\beta t})}}
\exp(2\beta t)\exp\left(-\frac{x^2 b^2\sigma}{2}\right)\,.
\ee
We now replace $x$ by $z$, considering it as a complex variable by 
analytic continuation. We consider again the observables 
$\cO_k(z)\equiv z^k,k=1,\ldots 4$ and $k=-1$; the integrals can be 
carried out analytically and yield
\be
\cO_1(z;t)=\frac{1}{bz}+b\sigma z\,,
\label{zbeta1}
\ee
\be
\cO_2(z;t)=3\sigma+b^2\sigma^2 z^2\,,
\label{z^2beta1}
\ee
\be
\cO_3(z;t)=\frac{3\sigma}{bz}+6 b\sigma^2 z
+b^3\sigma^3 z^3\,,
\label{z^3beta1}
\ee
\be
\cO_4(z;t)=15\sigma^2+ 10 b^2\sigma^3 z^2+b^4\sigma^4 z^4
\label{z^4beta1}
\ee
and
\be
\cO_{-1}(z;t)=\bra 1/z \ket_{\rho(t)}=\frac{1}{b\sigma z}\,.
\label{1/zbeta1}
\ee
As for $\beta=0$, the expressions for the odd powers are now meromorphic
functions of $z$, with simple poles at $z=0$. For the even powers we have
polynomials in $z$.

Consistency of these results with the earlier ones for $\beta\to0$ is
easily verified, using
\be
\lim_{\beta \to 0} b= \frac{1}{2t}\,,\quad \lim_{\beta \to 0} \sigma=2t\,.
\ee
We can now also consider the limit $t\to\infty$, using
\be
\lim_{t \to \infty} b= 0\,,\quad \lim_{t \to \infty}\sigma=\frac{1}{2\beta}\,.
\ee
For $k$ odd $\cO_k(z;t)$ grows exponentially in $t$, whereas for 
$k$ even 
\be
\lim_{t\to\infty}\cO_2(z;t)= 3\sigma= \frac{3}{2\beta}
\ee
and
\be
\lim_{t\to\infty}\cO_4(z;t)=15 \sigma^2= \frac{15}{4\beta^2}\,,
\ee
which are the correct expectation values.

\begin{figure}[ht]
\begin{center}
\includegraphics[width=0.48\columnwidth]{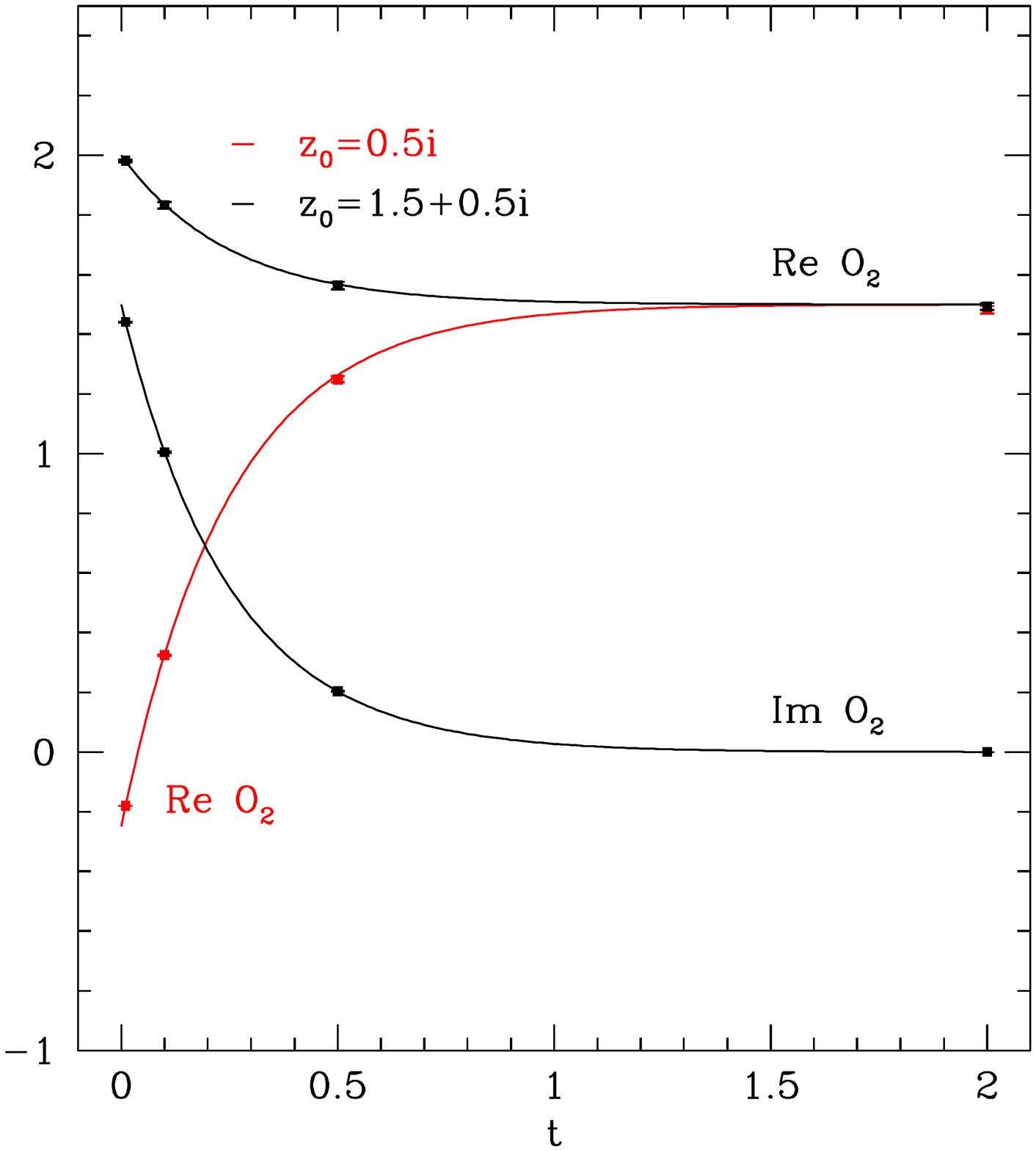}
\includegraphics[width=0.48\columnwidth]{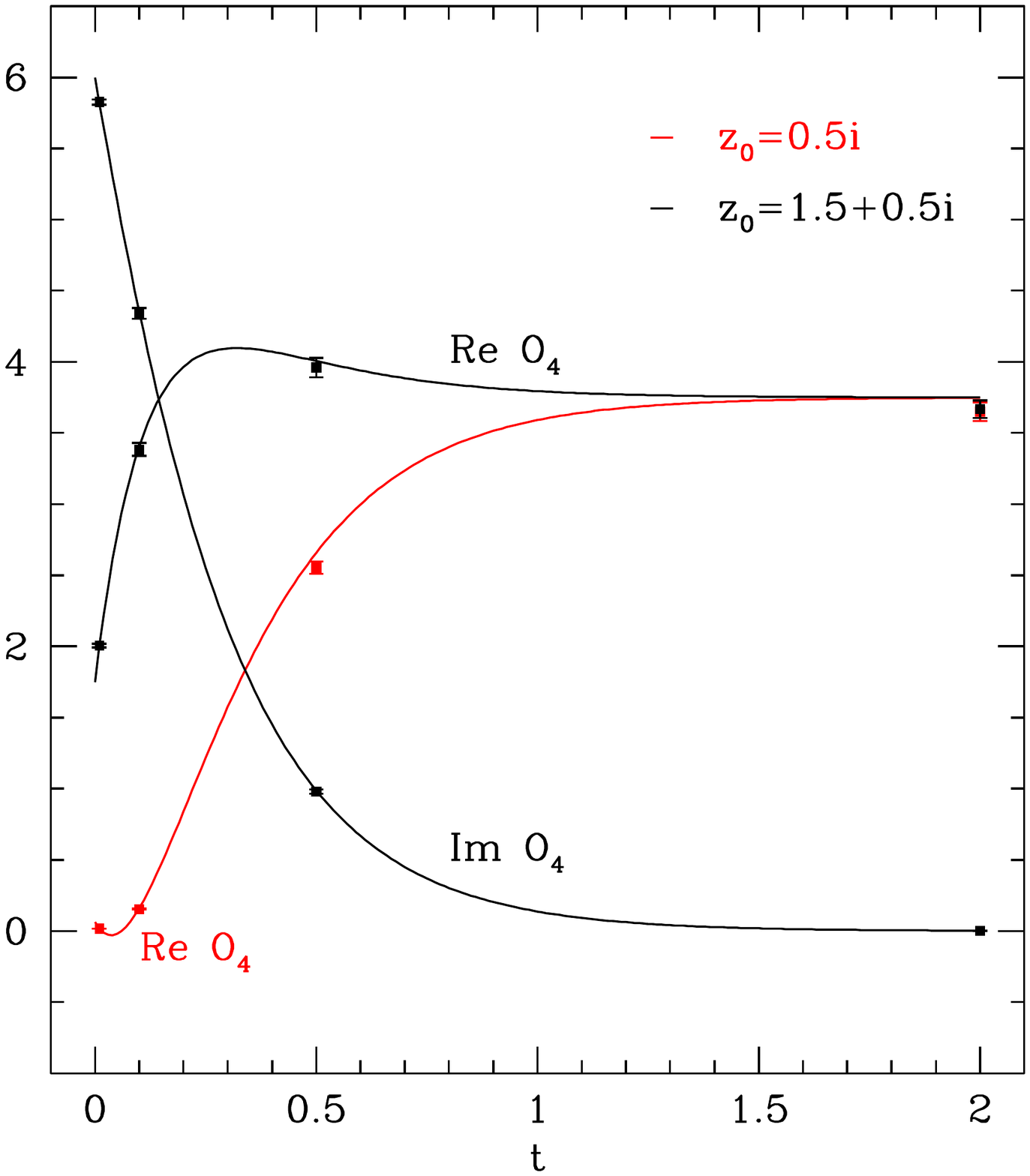}
\vglue-2cm
\caption{Comparison of $\cO_k(z;t)$ with $\bra {\cO}_k \ket_{t,z_0}$ for 
$k=2,4$ and $\beta=1$.}
\label{fpecomp1}
\end{center}
\end{figure}

We can again compare $\cO_k(z_0;t)$ and $\bra \cO_k \ket_{t,z_0}$ for 
finite times; qualitatively the situation is not different from the 
case $\beta=0$, see Fig. \ref{fpecomp1}, except that now the limit 
$t\to\infty$ can be considered. We find again agreement for $k$ even 
and disagreement for $k$ odd; for $t=2$ the expectation values of the 
even powers have almost reached the infinite time limit. 

As before, for $k$ odd there is strong disagreement between 
$\cO_k(z_0;t)$ (which has a finite limit for $t\to\infty$) and $\bra 
\cO_k \ket_{t,z_0}$ (which grows exponentially); this implies that the 
interpolating function has a nonzero slope, signaling a boundary term.

\subsection{General $n_p>0$}

For this case we can still give a rather complete analysis, even though
we do not have the benefit of the Mehler formula. We have
\be
L_c=(D_z+K)D_z=(D_z+\frac{n_p}{z}-2\beta z) D_z\,;
\ee
so the even and odd subspaces are still invariant under $L_c$. For a 
holomorphic observable $\cO(z)$, given by a convergent power series
\be
\cO(z)=\sum_{n=0}^\infty a_n z^n
\ee
we find
\be
(L_c \cO)(z) =\sum_{n=0}^\infty  a_n(L_c z^n) \equiv \sum_{n=-1}^\infty
(L_c^T a)_n z^n ,.
\ee
The dual action on the coefficients is thus
\be
(L_c^T)_n=(n+2)(n+1+n_p) a_{n+2} -2\beta n a_n\,.
\ee
Looking for eigenvalues of $L_c^T$ we find
\be
(n+2)(n+1+n_p)a_{n+2}-2\beta n a_n=\lambda a_n\,;
\ee
writing
\be
\lambda=2\beta k,\quad k\in \Z\,,
\ee
this can be rewritten as an upward recursion
\be
a_{n+2}=\frac{2\beta(n+k)}{(n+2)(n+1+n_p)} a_n\,.
\label{uprecurs0}
\ee
There are two choices to start the recursion:\\
(a) at $n=0$\\
(b) at $n=1-n_p$\\
and the recursion will stop at $n=-k$. So for fixed $k$ only finitely many
$a_n$ will be different from $0$ in both cases.

The semigroup $\exp(tL_c)$ applied to $z^k, k= 1-n_p,\ldots,1,2,\ldots$ will
thus be given by a polynomial in $z$ and $1/z$. We give a simple example:
\be
\cO_2(z;t)=\frac{1+n_p}{2\beta}\left(1-\e^{-4\beta t}\right)
+z^2\e^{-4\beta t}\,,
\ee
which generalizes (\ref{z^2beta1}) to general $n_p$. We now distinguish 
two cases:

\begin{itemize}
\item $n_p$ even:

Let $0<n_p=2\ell$, $\ell$ integer. The eigenfunctions of $L_c$ are in
case (a) even polynomials in $z$ and in case (b) odd polynomials in
$z$ and $1/z$, i.e. rational functions.

For the choice (a) the eigenvalues are nonpositive, corresponding to
$k=0,-1,-2, \ldots$ and the semigroup $\exp(tL_c)$ applied to
$z^{2\ell}, \ell=0,1,2,\ldots$ will be given by even polynomials in
$z$.

For choice (b) there are positive and negative eigenvalues,
corresponding to $k=\ldots, -2,-1, 0,\ldots n_p-1$. The semigroup
$\exp(tL_c)$ applied to $z^{2\ell+1},\ell=-n_p\ldots, 0,1,2,\ldots$
will be given by odd polynomials in $z$ and $z^{-1}$, with largest
negative power $z^{1-n_p}$. In particular $z^{1-n_p}$ is an
eigenfunction with the positive eigenvalue $2\beta(n_p-1)$.

\item $n_p$ odd:

Let $0<n_p=2\ell+1$, $\ell$ integer.

Again for choice (a) the eigenfunctions of $L_c$ are even polynomials in 
$z$. For (b) we obtain even polynomials in $z$ and $1/z$. But the 
linear space of the latter contains the polynomials in $z$ arising from 
case (a). There is only a finite ($\ell-$)dimensional space of 
polynomials in $1/z$ with the highest negative power being $z^{-2\ell}$, 
containing all the eigenvectors with positive eigenvalues. There are no 
odd eigenfunctions.

\end{itemize}

\section{The general case: $n_p>0$, $\beta>0$, $z_p\neq~0$}
\label{complex}

\subsection{Analytic considerations}

This case can no longer be solved analytically, whether $z_p$ is real 
or not. It was studied numerically in \cite{pole}. Here we discuss 
some mathematical subtleties arising in this case.

First we want to formulate a mathematical conjecture that might seem 
plausible, but is in general not correct:

{\bf Conjecture 1:} {\em Let $K(z)$ be meromorphic in $\C$, holomorphic 
in a domain $G\subset\C$ and $\cO(z)$ also holomorphic in $G$. Then 
there is a solution to the initial value problem (\ref{cauchy}), which is 
jointly holomorphic in $(t,z)$ for $z$ in any simply connected subset of $G$ 
and $t$ in a neighborhood of $\R_+=\{t|t>0\}$.}

But conjecture 1 is wrong even for the case of $K$ being an entire 
function, in fact even for $K=0$. The following counterexample is due to
Sophie von Kowalevsky \cite{svk}:

{\bf Counterexample:} Let $K=0$, i.~e. $L_c=D_z^2$ and 
$\cO(z)=1/(1+z^2)$. Then $\cO(z;t)$ is not analytic in $(t,z)$ at 
$(0,0)$.

{\em Proof.}-- $\cO(z;t)$ is given, using the heat kernel, by
\be
\cO(z;t)=\frac{1}{\sqrt{4\pi t}} \int_{-\infty}^{\infty} dy
\exp\left(-\frac{(z-y)^2}{4t} \right) (y^2+1)^{-2}\,.
\ee
A closed analytic form of this could be given in terms of error 
functions, but it is not needed. The nonanalyticity can be seen either 
by looking at a power series ansatz in $t$ and $z$ for the solution, 
which diverges for any $t\neq 0$, or by looking at the result for $z=0$, 
which is
\be
\sqrt{\frac{\pi}{4t}} \exp\left(\frac{1}{4t}\right) {\rm
Erfc}(1/\sqrt{4t})\,.
\ee
This is clearly not analytic in $t$ at $t=0$. \hfill$\square$

On the other hand we can express $\cO(z;t)$, using Fourier transformation, 
as
\be
\cO(z;t)= \frac{1}{2\pi}\int_{-\infty}^{\infty} dk \exp(-tk^2
-|k|)\exp(ikz)\,
\ee
showing that for any $t\in\C$ with ${\rm Re}\,t>0$, $\cO(z;t)$ is an entire 
function of $z$. 

Returning to our one-pole model, $-L_c$ is still conjugate to the Hamiltonian 
$H_{FP}$:
\be
H_{FP}=-(D_x+\frac{1}{2}K)(D_x-\frac{1}{2}K)=
-D_x^2-\frac{1}{2}S''+\frac{1}{4}(S')^2\,,
\ee
via the similarity transformation (\ref{simil1}). Inserting
\be
S'=-\frac{n_p}{x}+2\beta (x+z_p)\,;\quad S''=\frac{n_p}{x^2}+2\beta\,
\ee
the Hamiltonian becomes
\bea
H_{FP}&=-D_x^2+\frac{n_p}{4x^2}(n_p-2)+\beta^2(x+z_p)^2-
\frac{\beta n_p z_p}{x}-\beta (n_p+1)
\notag\\&=H_{even}+H_{odd}\,.
\label{h-even-odd}
\eea
with
\be
H_{even}=-D_x^2+\frac{n_p(n_p-2)}{4x^2}+\beta^2 x^2+\beta^2 z_p^2
-\beta (n_p+1)
\ee
and
\be
H_{odd}=-\frac{\beta n_p z_p}{x}+2\beta^2 x z_p
\ee

For nonreal $z_p$ this is a non-Hermitian operator and we do not know 
much about its spectrum; in fact to give the term a precise meaning we 
would first have specify the space in which $H_{FP}$ operates. A simple 
choice is $L^2(|\rho(x)|dx)$.

But let us try to find the action of the semigroup $\exp(tL_c)$ on powers 
of $z$. We have
\be
L_c z^n= n(n+n_p-1) z^{n-2}-2\beta n z^n-2\beta n  z_p z^{n-1}
\ee
and hence the dual action on the Taylor coefficients $a_n$ of $\cO$ is 
\be
(L_c^T a)_n = (n+2)(n+1+n_p)a_{n+2}-2\beta n a_n-2\beta (n+1) a_{n+1}
z_p\,.
\label{recurs}
\ee
The $z_p$ term mixes the even and odd subspaces.

\begin{figure}[ht] 
\begin{center}
\includegraphics[width=0.48\columnwidth]{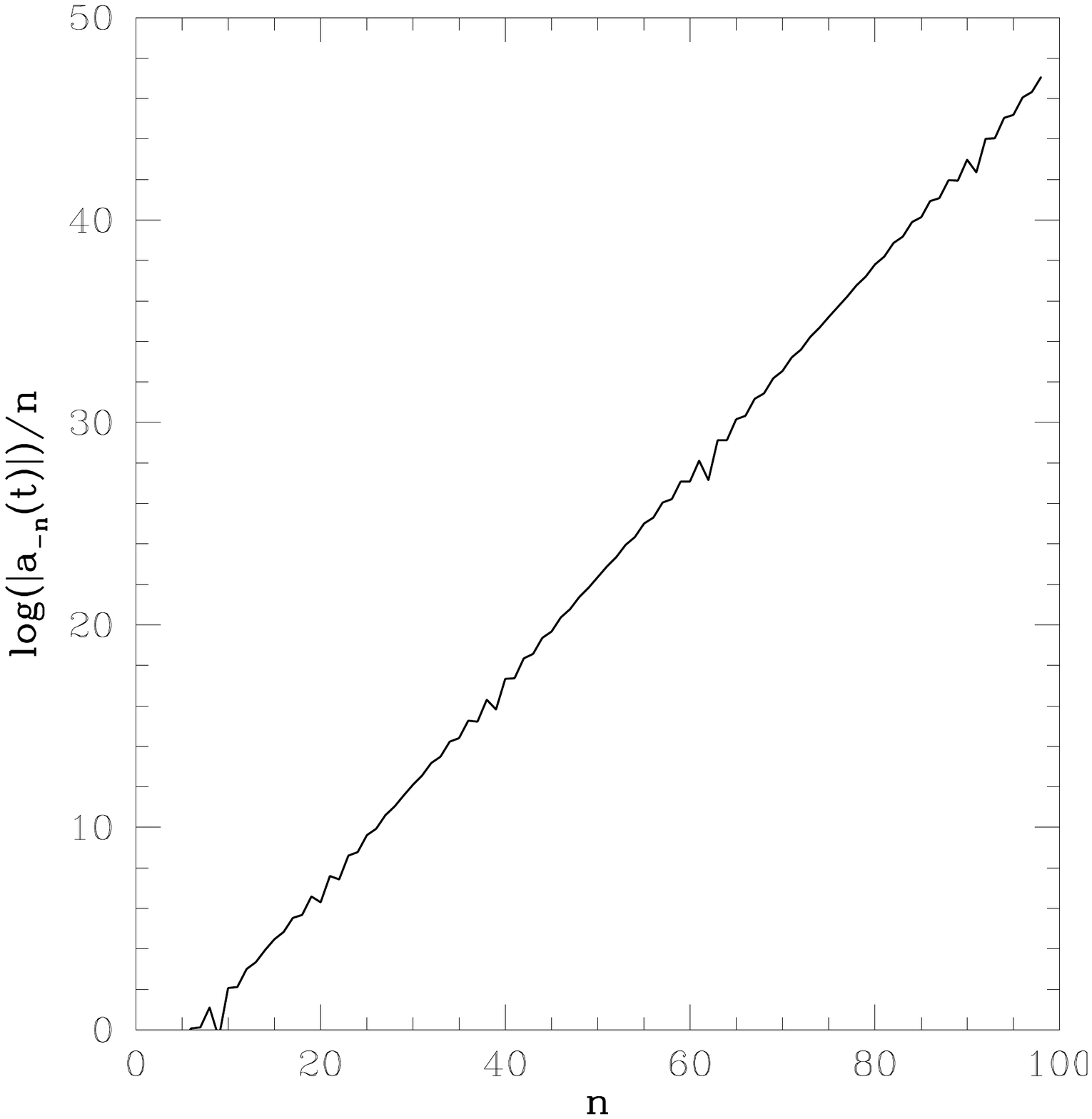}
\vglue-2cm
\caption{$\log(|a_{-n}|)/n$ for $n=1,\ldots, 100$ with parameters 
$\beta=1,\,z_p=i,\,n_p=2,\,t=0.5$.}
\label{laurent}
\end{center}
\end{figure}

Equation (\ref{recurs}) has the structure of a downward recursion; 
iterating this recursion will produce nonvanishing coefficients $a_n$ 
with $n$ arbitrarily large negative. Unfortunately this will lead to 
coefficients $a_n$ growing factorially for $n\to\infty$.

Formally, $(\exp(tL_c)\cO)(z)$ will be given by a Laurent series
\be
\left(\exp(tL_c)\cO_k\right)(z)= \sum_{n=-\infty}^k a_{n}(t)z^n\,, 
\ee
with the coefficients produced by exponentiating the recursion (\ref{recurs}). 
We show in Fig.~\ref{laurent} the expression
\be
\frac{\log|a_{-n}(t)|}{\log (n)}
\ee
for the observable $\cO_2(z)=z^2$ and the parameters given in the caption. 
It is seen clearly that there is a linear increase, indicating that
$|a_{-n}(t)|$ grows roughly like $n^{n/2}$, so the Laurent expansion 
diverges for any $z$.

We interpret this situation as follows: it is analogous to the one in 
the counterexample above: the semigroup $\exp(tL_c)$ applied to the 
powers $z^k$ is {\em not} analytic in $t$ at $t=0$, so it cannot be 
constructed via the exponential series. In the counterexample there was 
a simple way out of this dilemma: just avoid observables with poles. 
Here there is also a subspace of observables that produces a convergent 
Laurent series (see below). But for pure powers we cannot construct the 
semigroup by means of the exponential series. We expect that the 
solutions to the initial value problem (\ref{cauchy})
exist and are meromorphic in $z$, but in general nonanalytic in $t$ at 
$t=0$, just as in the counterexample above.

So the analyticity properties of the solution to the initial value problem 
for a general holomorphic observable, where it exists, are not as simple 
as conjecture 1 would suggest, undermining the formal justification of the 
CL method.

But as stated above, it is possible to find a linear subspace of 
observables that do not suffer from this disease; this subspace is 
obtained as a deformation of the even subspace for $z_p=0$. It can be 
obtained as the linear span of the eigenvectors of $L_c$ to nonpositive 
eigenvalues, which in turn are deformations of the eigenvectors obtained 
for $z_p$=0.

The dual action $L_c^T$ (\ref{recurs}) on the coefficients leads to 
the eigenvalue equation 
\be
(n+2)(n+1+n_p)a_{n+2}-2\beta n a_n-2\beta (n+1) a_{n+1}z_p=
\lambda a_n\,.
\ee  
This can again be rewritten as an upward recursion
\be
a_{n+2}=\frac{2\beta}{(n+2)(n+1+n_p)}\left[(n+k) a_n+(n+1)z_p\,
a_{n+1}\right]\,,
\label{uprecurs}
\ee
with
\be
\lambda=2\beta k\,,\quad k=0,-1,-2,\ldots.
\ee
The recursion has to start at $n=0$; the lowest coefficients are
\be
a_1=0\,,\quad a_2=\frac{\beta k}{1+n_p}a_0\,.
\label{initial}
\ee
For $z_p\neq 0$ the recursion no longer terminates, but it produces a 
sequence decaying roughly like $1/\Gamma(n/2)$, thus defining an entire 
function of $z$: it is not hard to prove by induction a bound of the 
form
\be
|a_n|<\frac{1}{\Gamma(n/2+1)}|a_0|\left(2|\beta|(1+|z_p|)\right)^{n/2}
\label{bound}
\ee
For observables from this space we do not expect any boundary terms and 
the CL method should work. 

We checked the correctness of the CL method numerically for the example 
$\cO_2$, the eigenvector of $L_c$ obtained for $k=-2$ from the upward 
recursion (\ref{uprecurs}). We ran the recursion up to $n=50$, when the 
coefficients $a_n$ are below $10^{-26}$. In Table~\ref{superobs} we 
compare the CL results with those of the $L_c$ evolutions. The 
parameters are given in the caption.

\begin{table}[ht]
\begin{center}
\begin{tabular}{|r| r | r|}
\hline
$t$ & $\bra \cO_2\ket_{t,z_0} $ &exact      \\
\hline
\hline
$0.00$ & 1.14289(0) &1.14289  \\
\hline
$0.01$ & 1.09798(12) & 1.09807 \\
\hline
$0.10$ & 0.7664(18) & 0.76610 \\                               
\hline                                                                        
$0.50$ & 0.1596(79) & 0.15467  \\
\hline
$1.00$ & -0.015(12) & 0.02093 \\
\hline
$2.00$ & -0.036(12) & 0.00038 \\                               
\hline 
\end{tabular}
\caption{CL results for the time evolution of the observable $\cO_2$, which is 
an eigenvector with eigenvalue $-4\beta$ for $n_p=2$, $\beta=1$, $z_p=i$;
the exact results are obtained as $\exp(-4\beta t)\cO_2(z_0)$ with $z_0=0.5i$.}
\label{superobs}
\end{center}
\end{table}

The table shows that the CL results are correct, possibly with a small 
truncation error for $t=2$.

It should be noted that the recursion (\ref{uprecurs}) works for any 
$k\in\R$, leading to an entire function of $z$. Of course $k>0$ should 
be excluded because it does not lead to convergence fore $t\to\infty$. 
But does this mean that the whole positive real axis belongs to the 
spectrum? This question is not well posed without specifying the space 
(Hilbert space, Banach space or a more general topological vector 
space) in which the problem is posed.

In a slightly different way, a subspace of entire functions invariant 
under $L_c$ is obtained by forming linear combinations of observables 
for various values of $k$. This means that the second condition of 
(\ref{initial}) is no longer enforced, but the condition $a_1=0$ still 
holds. The invariance under $L_c$ requires that $a_3$ is given as
\be
a_3=\frac{2\beta z_p}{2(n_p+2)} a_2\,;
\ee
to preserve this relation under $L_c$ enforces a similar linear 
relation between $a_5$ and $a_4$. Continuing this kind of reasoning, we 
learn that for any $\ell>0$ $a_{2\ell+2}$ is a fixed multiple of 
$a_{2\ell}$ with a factor that goes to zero at least linearly with 
$z_p$. The coefficients still obey the bound (\ref{bound}), so this 
defines the subspace $\cV_+$ of entire functions invariant under $L_c$; 
the elements of this subspace will not give rise to boundary terms and 
the CL process produces correct results for them.

It would be nice to find a similar invariant subspace, consisting of 
functions holomorphic in $\C\setminus \{0\}$, and which reduces to the 
odd subspace for $z_p=0$. We could not find such a space because of the 
the convergence problems of the Laurent expansion discussed above. But 
it is clear that for observables $\cO\notin \cV_+$ the formal 
justification of the CL method fails and boundary problems are to be 
expected..

\subsection{Direct numerical estimation of the boundary term at $\tau=0$} 

Even though we cannot solve the evolution of holomorphic observables 
in the general case, we can still study directly the boundary term 
$B_\delta$ [see (\ref{boundterm})] at the pole numerically. Applying 
integration by parts in the form of Gauss's and Green's theorems twice 
to (\ref{full_bt}) one finds 
\be
B_\delta= -\oint_{\partial G_\delta} \vec K \cdot \vec n \,
P_{z_0}(x,y;t)\cO(x+iy)ds+o(\delta)\,,
\ee
with $\vec n $ the outer normal to $\partial G_\delta$ and $\vec 
K=(K_x,K_y)$; see \cite{pole} for details. The terms lumped together in 
$o(\delta)$ can be seen to go to zero for $\delta\to 0$ in the same way 
we saw that in equilibrium $B_\delta\to 0$. Omitting further 
contributions which are $o(\delta)$ and omitting multiplicative 
constants, we can replace the observable $\cO$ by its value at the pole, 
replace this by $1$ and replace $K$ by $1/(z-z_p)$

So we are reduced to considering finally
\be 
B_\delta\equiv \oint_{\partial G_\delta} 
\frac{(x-x_p)^2-(y-y_p)^2}{r^3}P_{z_0}(x,y;t)ds\,.
\label{simple_bt}
\ee
with $r=\sqrt{(x-x_p)^2+(y-y_p)^2}$. [In our reasoning we assumed 
$\cO(z_p)\neq 0 $. If $\cO(z_p)=0$, one has to consider higher 
derivatives  of $\cO$, which cannot all vanish, but this would lead too 
far afield.]

To estimate (\ref{simple_bt}) numerically, we consider a sequence of rings
around the pole, given by
\be
\delta(1-\eta) <|z-z_p|<\delta(1+\eta)\,,\quad
\delta=0.08\,,\;0.04\,,\;0.02\,,\;0.01\,;
\ee 
with $\eta=0.1$ and $\eta=0.2$\;. We chose $z_p=-0.5i$ and the starting 
point $z_0=0.5i$. The first thing to note is that the CL process needs a 
certain amount of time $t_{min}$ before it reaches the line $y=y_p$. 
Since the motion in the $y$ direction is deterministic, we can estimate 
this minimal time by moving along the $y$ axis and find
\be
t_{min}\approx 0.23\,.
\ee
Secondly, it is difficult to achieve sufficient statistics near the 
pole, because the probability density $P$ vanishes linearly at the pole, 
where $|\vec K\cdot \vec n|$ is maximal, creating an ``overlap problem''. 
We ran $10^7$ independent trajectories, but still obtained only about 
50 hits for $\delta=0.01,\, \eta=0.1$ and $t=0.24$ and $0.25$, and even 
fewer for other values of $t$.

Therefore our results are not precise enough for a reliable extra\-polation 
to $\delta=0$. But we think it is fair to estimate $B_0\approx 0.15$ for 
$t=0.24$ and $t=0.25$.

\begin{figure}[ht]
\begin{center}
\includegraphics[width=0.7\columnwidth]{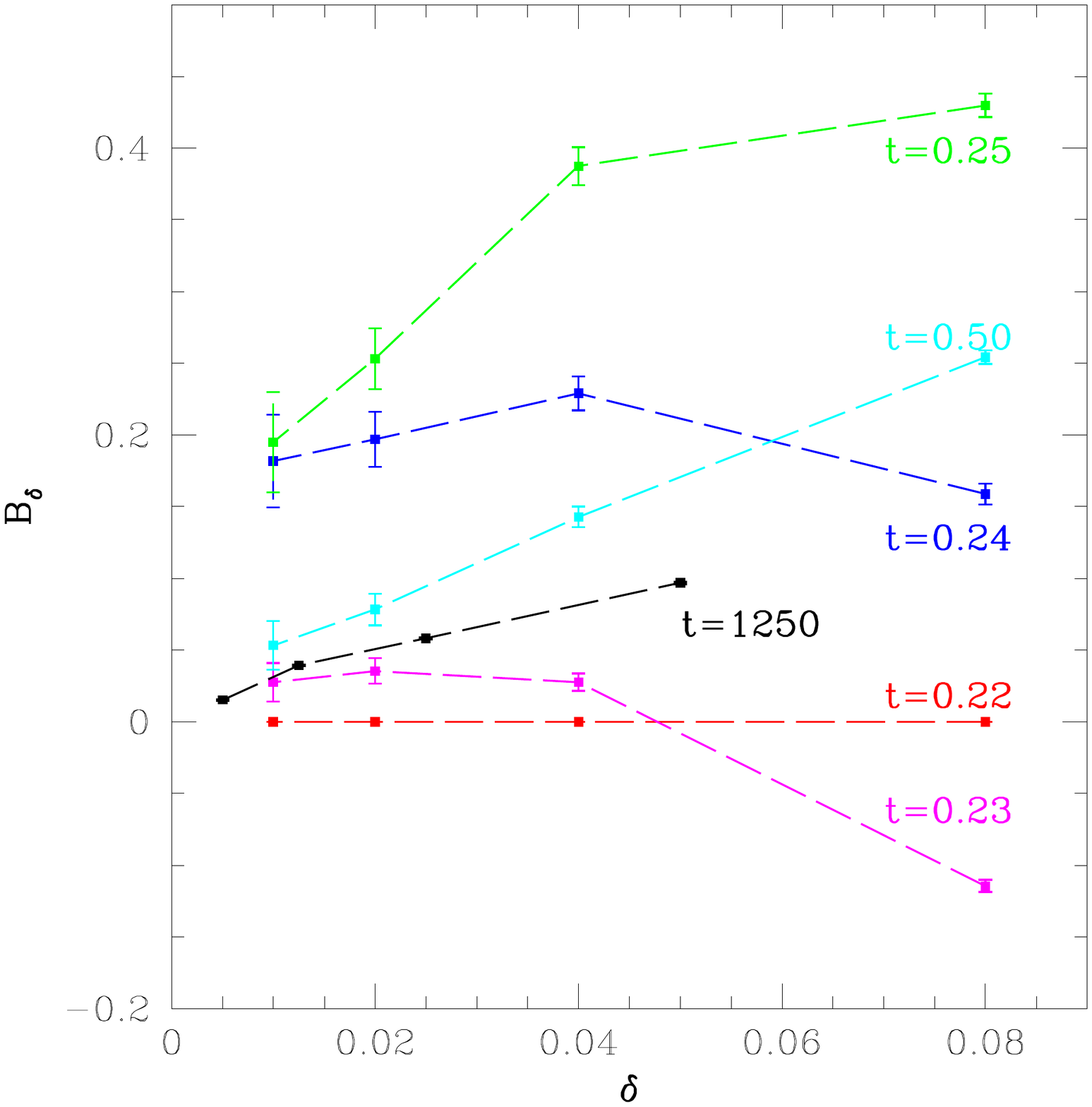}
\vglue-2cm
\caption{Numerical estimates of the boundary term $B_\delta$ for $\eta=0.1$
\;.}
\label{bc}
\end{center}
\end{figure}

The vanishing of $B_\delta$ for $t\le 0.22$ explained above is confirmed by 
the simulation. In Fig.~\ref{bc} we present the results for the Langevin times

\be
t=0.22\,,\;0.23\,,\;0.24\,,\;0.25\,,\;0.5\,
\ee
and for comparison we include also equilibrium data ($t=1250$). 

So the numerics indicate that beginning at $\approx 0.23$ indeed 
\be
\lim_{\delta\to 0} B_\delta \neq 0\,,
\ee
decreasing again with increasing $t$ and vanishing in the long time 
(equilibrium) limit.

\section{Conclusions and open problems}

The first conclusion is that to find boundary terms at poles, it is not 
sufficient to look at the equilibrium distribution; it is necessary to 
study the short time evolution.

The second point is that quite likely the conjecture 1 (analyticity in 
$t$ at $t=0$) is in general not correct for the simple observables like 
powers of $z$. It does seem to hold, however, for a subspace of 
holomorphic observables; for the one-pole model such a subspace has 
been constructed in Sec.~\ref{complex}; unfortunately for lattice 
models it seems very difficult imitate this construction.
 
Finally, for the pure pole model we established explicitly the 
existence of boundary terms at the pole by analyzing the short time 
evolution. Since in the vicinity of the pole the pole term always 
dominates, the pure pole model should give a good approximation of the 
CL process for more general models; thus we expect such boundary terms 
generally, provided the CL process comes arbitrarily close to the pole.

An open question concerns the analyticity properties that can be 
expected for the solutions of the general initial value problem 
(\ref{cauchy}), given the analyticity properties of the drift and the 
initial value $\cO(z;0)$. Since conjecture 1 failed, inspired by 
Kowalevsky's counterexample, we formulate something weaker:

{\bf Conjecture 2.} {\em Let $K(z)$ be meromorphic in $\C$, holomorphic 
in a domain $G\subset\C$ and $\cO(z)$ also holomorphic in $G$. Then 
there is a solution $\cO(z;t)$ to (\ref{cauchy}) which for $t>0$ is
holomorphic in $z$ for $z$ in any simply connected subset of $G$.}

{\em Remark.} Conjcture 2 of course implies that $\cO(z;t)$ can only have
isolated singularities at the poles of $K(z)$; these may be poles or 
essential singularities, but could also be branch points.
 
%

Conjecture 2 has implicitly been assumed to be true for instance in 
\cite{pole}. It certainly would be worth knowing if it can be converted 
into a theorem.

Our experience in the previous section unfortunately suggests the 
following:

{\bf Conjecture 3.} {\em Let $K(z)$ be meromorphic in $\C$, holomorphic
in a domain $G\subset\C$ and $\cO(z)$ also holomorphic in $G$. Then
``generically'' there is no solution $\cO(z;t)$ to (\ref{cauchy}) that is
holomorphic jointly in $(t,z)$ for $z$ in simply connected subsets of   
$G$ and $t$ in a neighborhood of $\R_+$.} 

The trouble is the lack of analyticity in $t$ at $t=0$, just as found in 
the example above. Of course the term ``generically'' is a bit vague; in 
our model we found that for conjecture 3 holds for $z_p\ne 0$ and 
$\cO\notin \cV_+$, but not for $z_p=0$ and not for $z_p\neq 0$, 
$\cO\in\cV$.

\vskip5mm
{\em Acknowledgments:} I would like tho thank Gert Aarts, Manuel Scherzer,
D\'enes Sexty and Nucu Stamatescu for the long and fruitful collaboration
on the CL method. Clearly this note is an outgrowth of that collaboration.
I am also grateful to Denes Sexty for useful comments on this manuscript.   

\appendix
\section{Importance of using the correct function space}

We want to highlight a subtlety of defining the semigroup $\exp(tL_c)$ or 
equivalently the initial value problem (\ref{cauchy})
by looking at a simple example. What we said about the spectrum is 
equally valid for the semigroup: without fixing the space in which we 
search for solutions, the initial value problem is not well posed.
 
As an example consider the observable $\cO_{-1}(z;0)=1/z$ for the pure 
pole model with $\beta=0\,, z_p=0\,,n_p=2$.

In (\ref{1/zfull}) we gave the solution as 
\be 
\cO_ {-1}(z;t)=1/z\,. 
\ee 
But there is a second solution: 
\be \cO_{-1}(z;t)=\frac{1}{z}{\rm Erf}\left(\frac{z}{2\sqrt{t}}\right)\,. 
\label{1/zhalf} 
\ee 

The difference lies in the analyticity properties in $t$: while for any 
fixed $t\in \R_+$ (\ref{1/zhalf}) is {\em holo}morphic in $z$ in the whole 
complex plane $\C$, it has an essential singularity in $t$ at $t=0$ for 
any fixed $z$. Furthermore, for $|{\rm Re}\,z| <|{\rm Im}\,z|$, the 
limit $t\to 0$ does not exist, so in this domain (\ref{1/zhalf}) does 
not solve the initial value problem.

Since the CL does not avoid the region $|{\rm Re}\,z| <|{\rm Im}\,z|$, 
for the solution (\ref{1/zhalf}) the argument for correctness fails. On 
the other hand, the {\em real} Langevin process, occurring for real 
starting points, never moves into the dangerous region and reproduces 
the second solution for $\rho(x)$ restricted to the half line $x\ge 0$.

On the other hand, as noted before, the solution (\ref{1/zfull}) is {\em 
mero}morphic in $z$ for fixed $t$, but {\rm holo}morphic in $t$ for 
fixed $z$, and solves the initial value problem correctly; here it is 
the pole at the origin that invalidates the formal argument by giving
rise to a boundary term.

So it is essential to specify at least the analyticity domains of the 
solutions when attempting to construct the solution of the initial value 
problem (\ref{cauchy}). The real and complex cases demand different spaces.

\end{document}